\documentclass[english,aps,onecolumn,superscriptaddress,showpacs]{revtex4}
\usepackage[T1]{fontenc}
\usepackage[latin9]{inputenc}
\usepackage{textcomp}
\usepackage{amsmath}
\makeatletter
\usepackage{epsfig}
\usepackage{changebar}
\usepackage{epsfig}
\usepackage{subfigure}
\usepackage{rotating}
\usepackage{amsfonts}
\newlength {\oldtextheight}
\newlength {\oldheadsep}
\psfigdriver{dvips}

\usepackage{babel}
\makeatother

\begin{document}
\title{Non-stationary aging dynamics in ant societies}
\author{Paolo Sibani and Simon Christiansen}
\email[]{paolo.sibani@ifk.sdu.dk}
\affiliation{
Institut for Fysik og Kemi, SDU, DK5230 Odense M, Denmark}
\pacs{87.23.Ge, 05.40.-a,02.50.Ng}
\date{\today}

\begin{abstract}  
In recent experiments by Richardson et al. ((2010), PLoS ONE 5(3): e9621. doi:10.1371/
journal.pone.0009621)
 ant motion out of the nest is shown to be  a non-stationary process 
intriguingly  similar   to the dynamics encountered in   \emph{physical aging}   of  glassy systems.
Specifically, exit events  can be described as 
a    Poisson process in logarithmic time, or, for short,  a log-Poisson process.
 Nouvellet et al.(J. Theor. Biol. 266, 573. (2010)) criticized these conclusions
and  performed new experiments where the exit process could  more simply be described
by standard  Poisson statistics.   In their reply, (J Theor. Biol.  269,  356-358 (2011))
Richardson et al.  stressed that the two sets of experiments 
were performed under very different conditions and claimed that this was the  likely source
of the discrepancy.   Ignoring any  technical issues which are part of the above
discussion, the focal point of this work  is to ascertain whether or not both  log-Poisson and Poisson statistics
  are possible in an ant society under different 
external conditions.
To this end, a model  is introduced where 
interacting  ants  move in a stochastic fashion from one site to a neighboring 
site  on a finite 2D lattice.  
The probability of each  move is determined by the ensuing changes of a utility function
which  is a sum of pairwise interactions
between ants, weighted by distance. Depending on how the interactions are
defined and on a control parameter dubbed  `degree of stochasticity' (DS), the dynamics 
 either quickly converges to a stationary state,
where  movements are a standard  Poisson process, or may enter a  non-stationary
  regime, where  exits can be described as  suggested  by Richardson et al.
Other aspects of the model behavior are also discussed, i.e. the time dependence of 
the average value of the utility function, and the statistics of spatial rearrangements happening 
anywhere in the system. Finally, we discuss the role of  record events and their statistics 
in the context of ant societies and suggest the possibility that a transition from non-stationary
to stationary dynamics can be triggered experimentally.
\end{abstract}
\maketitle

\section{Introduction}
\label{introduction}  
In their  study of  motion in  Temnothorax  albipennis ant colonies,   
 Richardson et al. \cite{Richardson10}
 found  that  ants leave their nest 
 at a decelerating  rate, and that the sequence of   exit times 
can be modelled using a peculiar  Poisson process, for short log-Poisson process, also known
from   record dynamics in  glassy systems\cite{Sibani93a,Sibani03,Anderson04}.
Subsequently \cite{Richardson10a},  the same group  experimentally demonstrated that social interactions 
are a prerequisite for  log-Poisson behavior.
Motivated by  Ref.~\cite{Richardson10}, Nouvellet et al.\cite{Nouvellet10}   
performed different experiments on the motion of Pharaoh's ants within their nesting area. The
results could be described by  a `null' model, 
where ants act independently  
and where the activity is  described by a  standard Poisson process, as
 fitting for a stationary and  time homogeneous situation.   
  In their  reply~\cite{Richardson11}, Richardson et al.  stressed that the two sets of experiments 
were performed under  different conditions and claimed that this would lead to 
different  event statistics. For completeness, we also  mention 
that    Nouvellet et al. recently replied~\cite{Nouvellet11} to  the comment of Richardson et al..

The discussion between the two groups  raises two  important theoretical issues.
One is whether  the  observed differences 
simply  reflect  the ability of the same system (or same type of system)
to display either behavior depending on some external condition
or parameter, as is the case in  glassy systems
at high and low  temperatures.
The second question regards the possible  interpretation of records
 in the context of ant dynamics, and
more generally, in the context of the dynamics of social agents.
The  dynamics of the simple model discussed below suggests some  answers. 

In the model,  ants move in a probabilistic fashion within a toroidal lattice, where contiguous points
each  represent  contiguous  areas of  physical space. Allowed moves    
involve a single 
ant changing its position from  one lattice point to one of its  four neighbors. 
 Moves  are accepted  or rejected with  probabilities 
depending  on the ratio of the   change they entail in a utililty function $E$
to a parameter $T$  dubbed `degree of stochasticity' (DS).
Specifically,  the DS  gauges the willingness   of  ants to make   risky  choices
which decrease   the value of their  own (and hence the whole system' s) utility function. 
These choices are akin to  investing  capital for the sake of future gains. The lower the DS,
the less inclined   ants are  to do so, i.e. at zero DS  only  moves which increase 
$E$ are performed. In the opposite limit of very high $T$, moves are  accepted regardless of the sign and magnitude of 
the corresponding change in $E$.
From an algorithmic point of view, $E$ and $T$  play a role similar to the energy and temperature
in a Monte Carlo simulation of a physical system. More importantly, the model dynamics and
the dynamics  of glassy systems are similar, a property  which the chosen  notation is meant to
emphasize. Nevertheless,  no relation is implied  between the DS parameter $T$ and the 
physical temperature of the environment in which  ants move. 
 At low DS, and
 for certain types of interactions but not for others, 
  a non-stationary dynamical regime is found, 
in many respect similar to the  \emph{physical aging} regime of glassy systems.
 The experimental findings of Richardson et al.
 are reproduced in this non-stationary regime.
Again to stress the connection to glassy dynamics,  the latter regime will is also called `aging' regime.
 It should be kept in mind, however, that 
in discussing the properties of the model  `aging' only  refers to the gradual changes
in the collective dynamical  behavior of  the whole colony. Correspondingly, the `age' of the system $t_{\rm w}$ is simply the
time elapsed from the start of the dynamical process through which spatial structures 
emerge  starting from a completely 
structure-less  configuration.  In contrast, the model does not  contemplate the possibility
that single ants  could change their characteristic response to the same external 
stimuli. It  hence excludes, \emph{a priori},   any aging process affecting  individuals. 
  The model also features a stationary regime, 
where movements are  well described by a standard Poisson process, similar
to the findings  of  Nouvellet et al. We speculate at the end on how
these two  regimes can arise   in ant colonies and on whether it is possible to trigger a change from 
one to the other.

\section{Model description}
\label{model} 
 Ants  of $N$ different types move  on a toroidal  lattice     of size $M = L^2$, where
 $L$ is the  linear grid size. Here, a  `type'   is only defined
 in terms of interactions and no  other properties, e.g. size and color, are attached  to  it.
 The interactions determine, in turn, how ants move around.
 In the following,   lattice points are indexed  using typewriter order,  i.e. the natural sequential ordering  of  letters in
a  text  written in a western language.   A configuration
 is then specified by the number $n_{l,x}$ of individuals of type $l$ located at site $x$,
 i.e. by the $N\times M$  rectangular matrix ${\mathbf n}$.
The  interactions $J_{ij}$ between
a pair of  individuals of type $j$ and $i$ are ordered in the 
 $N\times N$ square matrix $J$. 
Without loss of generality (see below), the latter can be assumed to
 be   symmetric.
Denoting matrix transposition by a superscripted  dagger, ($^\dagger$), we construct 
the $M\times M$ matrix 
\begin{equation}
{\mathbf I} = {\mathbf n}^\dagger  {\mathbf J }  {\mathbf n}.
\end{equation}
Each  entry  $I_{xy}= \sum_{i,j=1}^N n_{ix} J_{ij}  n_{jy}$   represents the interaction  
of all individuals at site $x$ with all individuals at site $y$, irrespective 
of the Euclidean distance $d_{x,y}$ separating these two sites.
 To weigh  distance in,  we finally  introduce
a `damping' matrix $\mathbf D$,  with entries
$D_{x,y} = \exp(-\alpha d^2_{x,y} )$, where $\alpha $ is a non-negative constant.
Note that $D$ is symmetric and has diagonal elements equal to one.
 The  sum of all interactions  finally produces the utility function 
\begin{equation} 
E = \mbox{\rm trace} ({\mathbf I} {\mathbf D}) = \sum_{x=1}^M \sum_{y=1}^M I_{x,y}D_{y,x}.
\end{equation}  
Each element of the double sum represents the interaction of individuals located at positions $x$ and $y$,
duly modified by the damping factor $D_{y,x} $.
The change in utility  associated to the move of an individual  of type $l$  from site $f$(rom) to site $t$(o) is  
correspondingly given by
\begin{equation} 
dE_{l,t,f} = 2  \sum_{x=1}^M \sum_{i=1}^N (D_{x,t} - D_{x,f}) n_{i,x} J_{l,i} + 2 J_{l,l} ( 1 - D_{t,f}).
\label{energy_change}
\end{equation}
Were the matrix $J$  not  symmetric, the  term 
$2J_{l,i}$ in the  above  expression would be replaced by
$J_{l,i} + J_{j,l}$.   As anticipated,  symmetric interactions can therefore  be 
assumed from the outset with no loss  of generality.
Secondly, if the system only contains a single ant of type $l$, i.e. if $n_{i,x} = \delta_{i,l}\delta_{x,f}$,
 the  change in utility associated
to any move from $f$ to $t$ is always zero.  
Likewise  if the interactions are    independent  of distance, e.g. if $D$ has all
entries equal to one. In both cases,  motion is   purely diffusive.

In general, the dynamics is a  Markov chain generated by 
the Metropolis rule.  In the stationary state eventually reached  by the
dynamics, the average value of the utility function $E$   decreases 
with increasing DS, while in a physical system the average energy increases with temperature.
Each  update comprises the following steps:
 a position is randomly chosen
with equal probability among all those  available. A type  is then   chosen in the same fashion.
An  ant of the given type (if present) is assigned a candidate  move to a randomly chosen neighboring site. 
The move is accepted if it increases the 
 value of the utility function.  Moves decreasing the
utility function  are accepted with a probability which decays exponentially as a function
of the ratio $dE_{l,t,f}/T$ of the  change in $E$ to the 
value $T$ of the DS.   For $T=0$   the dynamics is  a greedy optimization
algorithm attempting to maximize the utility function.  
 In the  stationary distribution of the Markov
chain just defined, the 
probability for   configuration
 $k$   is, modulo a normalization constant,  equal  to $\exp(E(k)/T)$.
Whether the stationary distribution is within reach  strongly
depends on the DS,  as further discussed below.
Time evolution is gauged in terms of  Monte Carlo (MC) sweeps,  each sweep 
comprising  a number of queries equal to the number of individuals in the system.  
In the initial configuration, the number   of ants of each type which are
located at a site  is drawn, independently for each site and type,   from  a uniform  distribution 
between $0$  and   $10$.

To connect with   ant experiments, an arbitrary site on the grid is designated   
  `exit', and all movements which involve this particular site are recorded as 
`events'.  Depending  on the situation, we either consider the statistics  of 
 the waiting times $ t_k- t_{k-1}$ between consecutive events, 
or that of the corresponding log-waiting times $\ln(t_k) - \ln(t_{k-1})$.
 Secondly, we calculate  the 
time dependence of the average utility function  for a number of different situations. Thirdly and  finally,
we show that, in the aging regime,  the probability of `large' events 
(defined later in the text) occurring in an arbitrary  time  interval $[t_w,t]$
anywhere on the grid  scales 
as $\ln(t/t_w)$, modulo finite time corrections due to  
 our time variable being restricted to integer values.

\section{General considerations}
Depending on a number of choices detailed below, the model  either
quickly reaches  a stationary state or is  unable to  
do so  within numerically accessible time scales.
In the first case, the `exit' process is 
a standard Poisson process. In the  second, 
a Poisson process also describes the statistics  but 
with the logarithm of time replacing time in its  average.
A logarithmic time dependence also characterizes the statistics of 
spatial re-arrangements  occurring anywhere in the system. 
It does seems that the replacement $t \rightarrow \ln t$ restores the 
time-homogeneity of the model dynamics at low values of $T$.

Behind the decelerating nature of aging dynamics is 
the gradual \emph{entrenchment} of dynamical trajectories in more
long-lived metastable configurations\cite{Sibani03,Anderson04}. In   our model
ants which attract each other tend to cluster. Starting from a random distribution,
ever larger clusters get established on gradually fewer sites. As discussed
below, this in turn creates growing
dynamical barriers (e.g. empty sites) for ants which have not yet
joined a cluster.

To investigate  the issue further several choices of interactions
have been considered. Our first choice  is  a null model lacking  
any metastable configurations. Ants of the same kind  repel each other, 
while ants of different kinds attract each  other.
Correspondingly, the interaction matrix has diagonal and 
off-diagonal elements equal to $-1$ 
and   $1$ respectively. In this case, 
ants of the same type tend  to maximize their mutual distance
and  hence   spread out uniformly in space,  irrespective of type.
Sites end up either being  empty or being occupied by    
 ants of different  types. 
For \emph{all } $T$ values,  the  dynamics quickly converges   to a stationary state,
where  `exit'
events are  (nearly) a standard Poisson process, of the sort  
illustrated (for a different example)  in the left panel of Fig.\ref{corr_and_logP}. 
For all $T$ s, the average $E$ relaxes  in a way similar to 
 the high $T$  curves ($T=200$ and $500$) shown in the left panel of Fig. \ref{big_energy_plot}.
 
More interesting  is the case  where 
ants of the same type are indifferent to each other, while
ants  of different types  attract each other, 
possibly  with the exception of ants of type $1$ and $2$, which 
 repel each other.
Correspondingly, $J_{ii} = 0$ for all $i$, $J_{12} =J_{21}= -1$ and  $J_{ij} =1$ for    any other values of 
$i$ and $j$.
Here, the dynamical  behavior strongly depends  on the value   of the $T$: stationarity
is reached quickly at  sufficiently  high $T$ but is unachievable at low $T$.

Consider first  the case of   two  types of ants with attractive interactions.
 A  high values of $E$  (low $T$) equilibrium configuration  
has the two types grouped  into a  small  number of sites and, eventually, into a single site.
Starting from a random initial distribution, 
empty areas  gradually   form. Since ants belonging to two metastable cluster located
 at different sites can increase their utility function by 
 a merging process which requires   crossing growing
empty areas, correspondingly  growing  dynamical
barriers  between metastable configurations are present in the model. As shown in Fig. \ref{corr_and_logP}. this
 model version has Poisson and log-Poisson exit statistics at high, respectively low $T$. 

Consider now  four types, with pairs of type ($1-2$)   repelling each other,
and all  other pairs  attracting  each other.  
Spatially segregated domains  of type $1$ and $2$ must gradually   form, while   groups of ants of type
$3-4$, each centered on a particular site, form  within each of the two domains.
Besides empty space,  a  domain $d_2$ containing ants  of type $2$   presents
an additional  dynamical 
barrier for  ants of type $1$,  since it either must be  crossed   or circumvented 
in order for these  to join a domain $d_1$ located  on the opposite  side of $d_2$.
Also in this case, low $T$  aging behavior is found to characterize the dynamics.

\section{Data analysis}
In all our simulations, the damping parameter  is $\alpha$ is equal to $5$, 
i.e. interactions are strongly localized in space.
The  spatially averaged  density  of 
each type  of ant  is close to $1.8$. E.g.,  with four   types present, the system  
contains appr. $7.2$ ants per grid site.  Linear grid sizes $L=5,7$ and $9$
were investigated for four ant systems and seen to have  qualitatively similar
behaviors.  
Linear grid size $L=7$ was used for systems with  $2,3,4$ and $5$ ant types.
The result shown are for $L=7$ to emphasize that a large grid size is not
required to obtain aging behavior.

\begin{figure}
 $
\begin{array}{cc}
\includegraphics[width=0.45\linewidth]{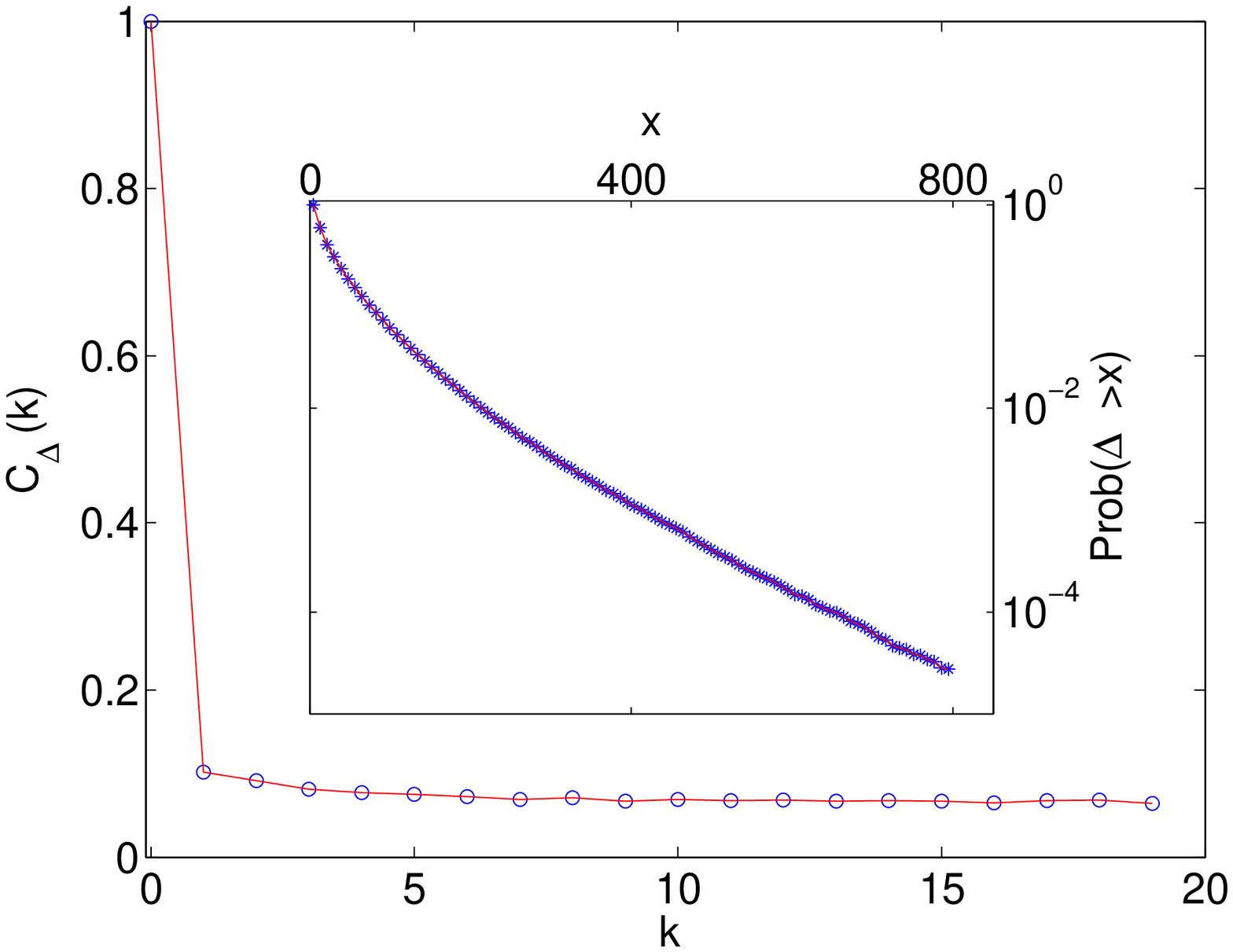}  & 
\includegraphics[width=0.45\linewidth]{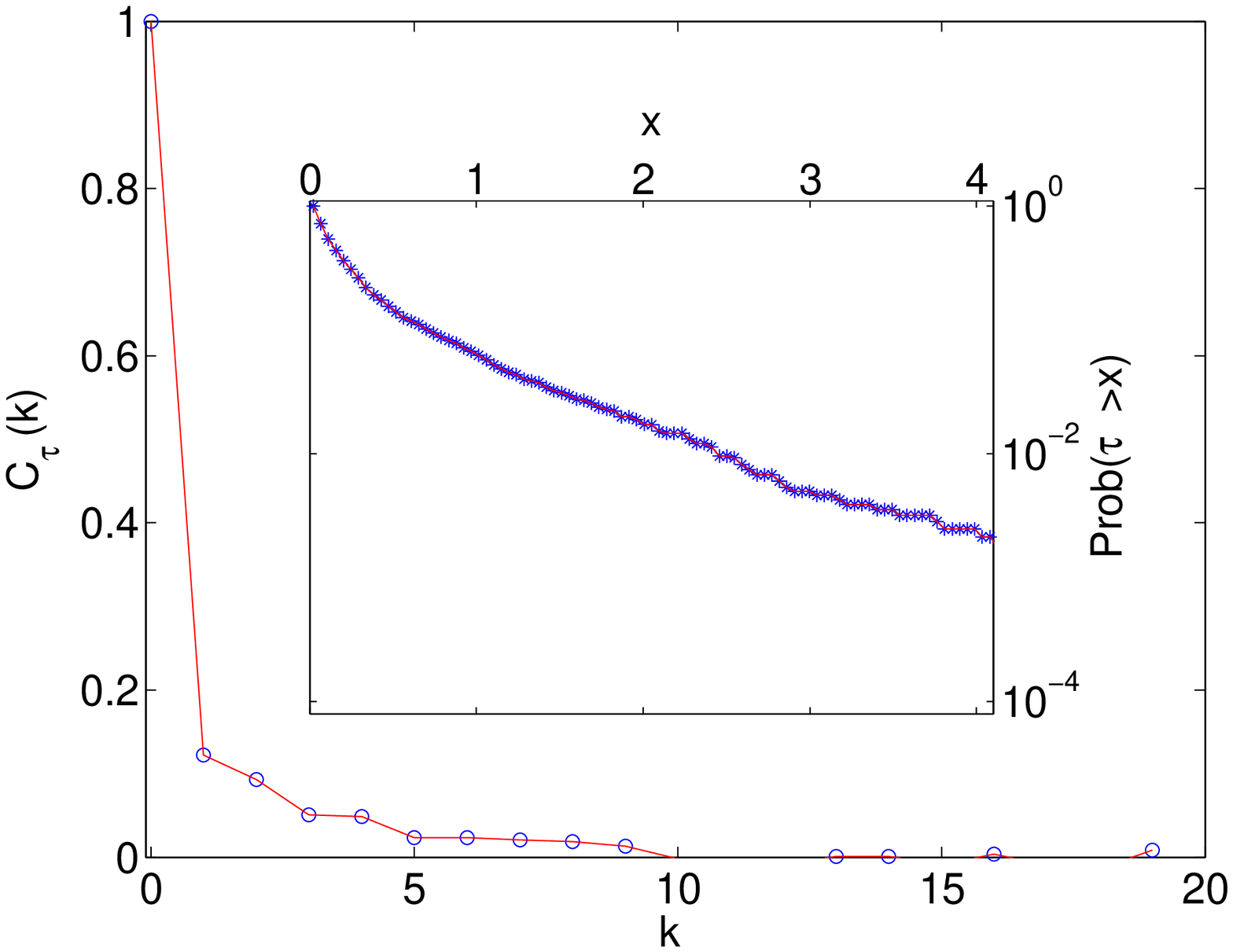}
\end{array}
$ 
\caption{(Color online)
Left: The correlation function $C_{\Delta}(k)$ for the waiting times
between  consecutive exit events, plotted  vs. $k$. The  insert shows the
cumulative distribution of the  waiting times, plotted on a logarithmic vertical scale.
The  system contains  two types  of ants moving on a grid of linear size $7$ with  DS
parameter is $T=50$.
Right: same as above, except that the DS value  is here  $T=5$ and that 
the correlation and cumulative distribution are calculated using the log-waiting times rasher than
the waiting times. 
}
\label{corr_and_logP} 
\end{figure}

Figure \ref{corr_and_logP} describes the exit statistics in a  system where  two types of ants are present.
At the $k$' th  sweep the program checks whether motion has occurred at the site dubbed `exit' and, if so,
registers the corresponding  time $t_k$. The   simulations, each running from
$t=5$ to $t=5 \cdot (1 + 10^5)$, are 
 repeated $100$ times   in order to improve the statistics.
 The DS values used in the simulation are  $T=50$ and $T=5$ in the left and  right panel, respectively.
The  data shown are statistically very different in spite of being graphically rather similar.
In the left panel  the waiting times, i.e. the  time differences $\Delta_k = t_k - t_{k-1}$
are analyzed with respect to their correlation  and
their distribution.
Since the    $t_k$' s are     integer rather than real numbers,  the exit  process can never    be truly Poissonian. 
We nevertheless  estimate  the normalized correlation function of the $\Delta_k$'s, averaged over 
$100$ independent runs.
For independent entries, the latter would equal   the  Kronecker's delta 
 $C_\Delta(k) = \delta_{k,0}$. We furthermore  estimate the probability that
 $\Delta >x$, as a function
of $x$. For a Poisson process this probability  decays exponentially in $x$. The correlation
and probabiliity distribution  are
plotted in the main figure and the insert, using  a  linear and a logarithmic ordinate, respectively.
We see that short waiting times, i.e. $\Delta_k$'s  of order one are over-represented 
 relative to the straight line representing  
the Poisson case. Secondly, the correlation decays to about $1/10$ in a single step,
but then lingers at that value. Taken together, these two feature  indicate that 
a  short waiting time is more likely followed by another short waiting time, i.e. that the motion
often stretches  over several  sweeps.

The right panel of the  figure shows data obtained   as just discussed, except that  logarithmic time differences 
$\tau_k = \ln(t_k) -\ln(t_{k-1})$ rather than  linear ones  are utilized. 
The correlation function $C_\tau(k)$ decays  quickly to zero, albeit
not in a single step, and the probability that $\tau > x$
is nearly exponential. Again,  short log-waiting times are over-represented in the distribution,
and since the correlation decays to near zero in $k=5$, they are likely 
to follow each other. Thus, also in this case   ant  at the 'exit'  
stretches beyond  a single sweep. In summary, banning the effect of our  time unit, the
sweep,  being too short relative to the de-correlation time of ant motion,,
  $T=50$ data are,  as expected  in a stationary regime,  well described 
by a Poisson process, while   
$T=5$ data are well described by  a log-Poisson process.

The values of the utility function and the DS  strongly affect  ant  motion:
The first  gauges
the well-being of the ants  or the degree to which a desirable aim
is reached by the ant society.
  Its value  could possibly be empirically quantified  for 
 agents which can be polled. The DS  measures the willingness to 
perform unpleasant moves, or to invest accumulated capital. As high DS value  signals indifference, 
 administering suitable drugs to the ants could possibly increase the DS.
   
The average value of the utility function  plotted vs. time  provides a good overall characterization 
of the model dynamics.
At low DS values,  the statistics of the  fluctuations of the utility function provides further 
insight on the nature of ant motion. We note that while exit events clearly reflect 
what happens at a given site, changes in $E$ can occur due to motion anywhere in the system.
At the level of a single trajectory, $E$  appears  most of the time to be a  constant.
  I.e. its fluctuations, defined as  differences between consecutive sweeps,
are most of the time equal to zero.  Correspondingly, most of the time, ants  hardly  move 
outside the area represented by a single grid point.
  Occasionally, a \emph{negative} fluctuation is quickly  followed by one or more   \emph{positive}  fluctuations.
Such events indicate   motion across a barrier from one metastable configuration to another, 
the latter usually   having a higher value of $E$.
Positive fluctuations larger than he threshold value $\Delta E_{\rm tr} = 0.22$
are counted as large events, irrespective of where they occur in the system. 
The  threshold is  chosen to filter our small fluctuations which could  
easily be reversed.   The  overall shape of the statistics of large events  is, within bounds,    insensitive to the specific choice
of threshold. 

Considering that, at sufficiently low $T$,  the average $E$ grows logarithmically in time, 
the typical number of large  fluctuations can be expected to do the same.
For  $k=0,1,\ldots 5$, consider observation  time intervals of the form $[t_w,\frac{3}{2}t_w]$ where  $t_w=800\quad  2^k$.
Increasing $k$ by one unit  doubles  the length of the interval, but  the difference between the
logarithm of the end-points remains in all cases   equal to  $\ln(3/2)$.
If the statistics of large events were only dependent on this  logarithmic difference, 
the probability $P_n(t_w,\frac{3}{2}t_w)$ that at least $n$ `large' events
occur in any of the  intervals  would be independent of $k$, and 
 probability functions obtained  for different values of $t_w$ would collapse
 on the same graph. This can never be exactly true: since  at most one event can 
 be registered   per MC  sweep,  the highest 
 number of events possible is limited  by 
the length of the observation interval. I.e. finite time corrections to the logarithmic law can be expected.

\begin{figure}
$
\begin{array}{cc}
\includegraphics[width=0.45\linewidth]{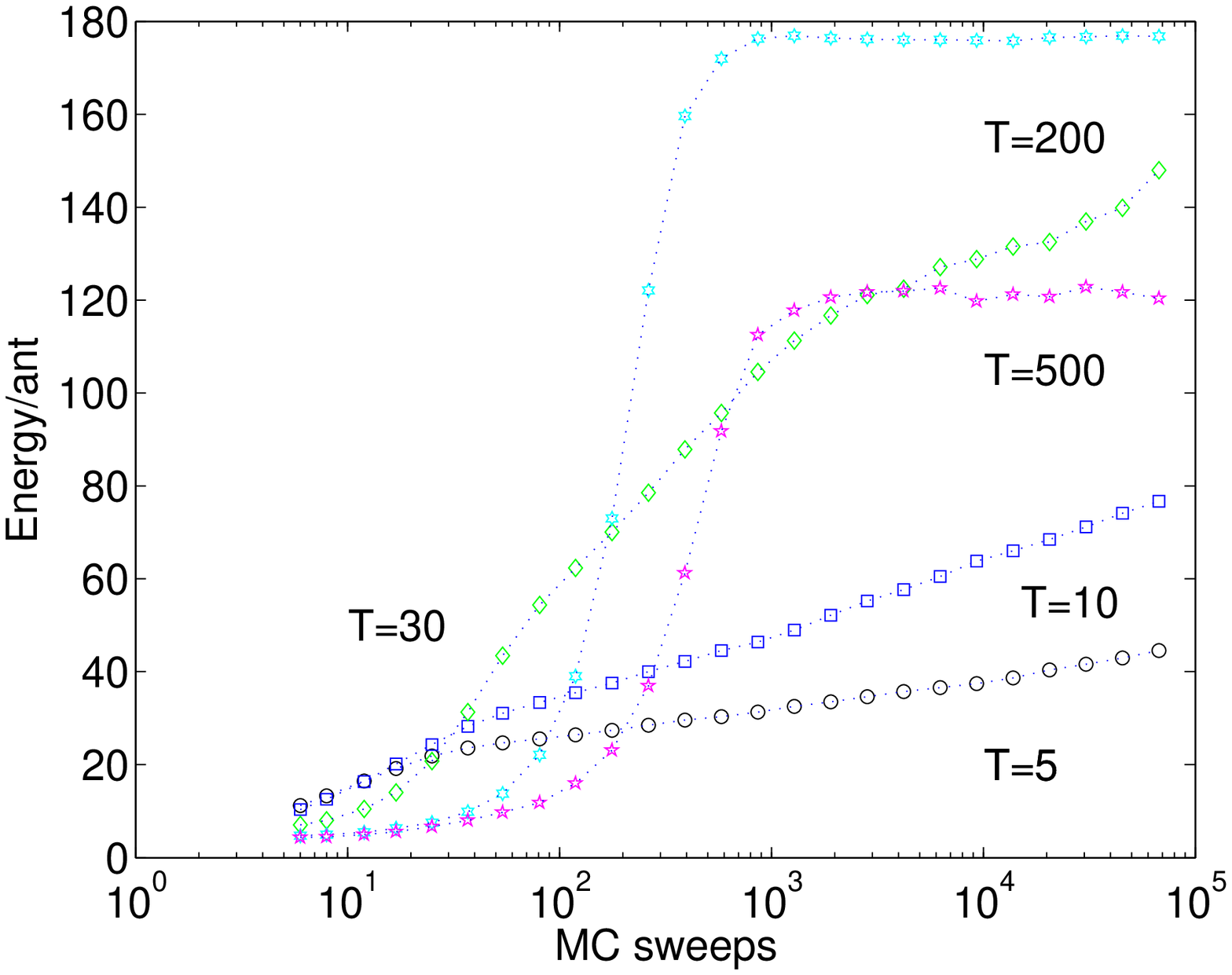} &
\includegraphics[width=0.45\linewidth]{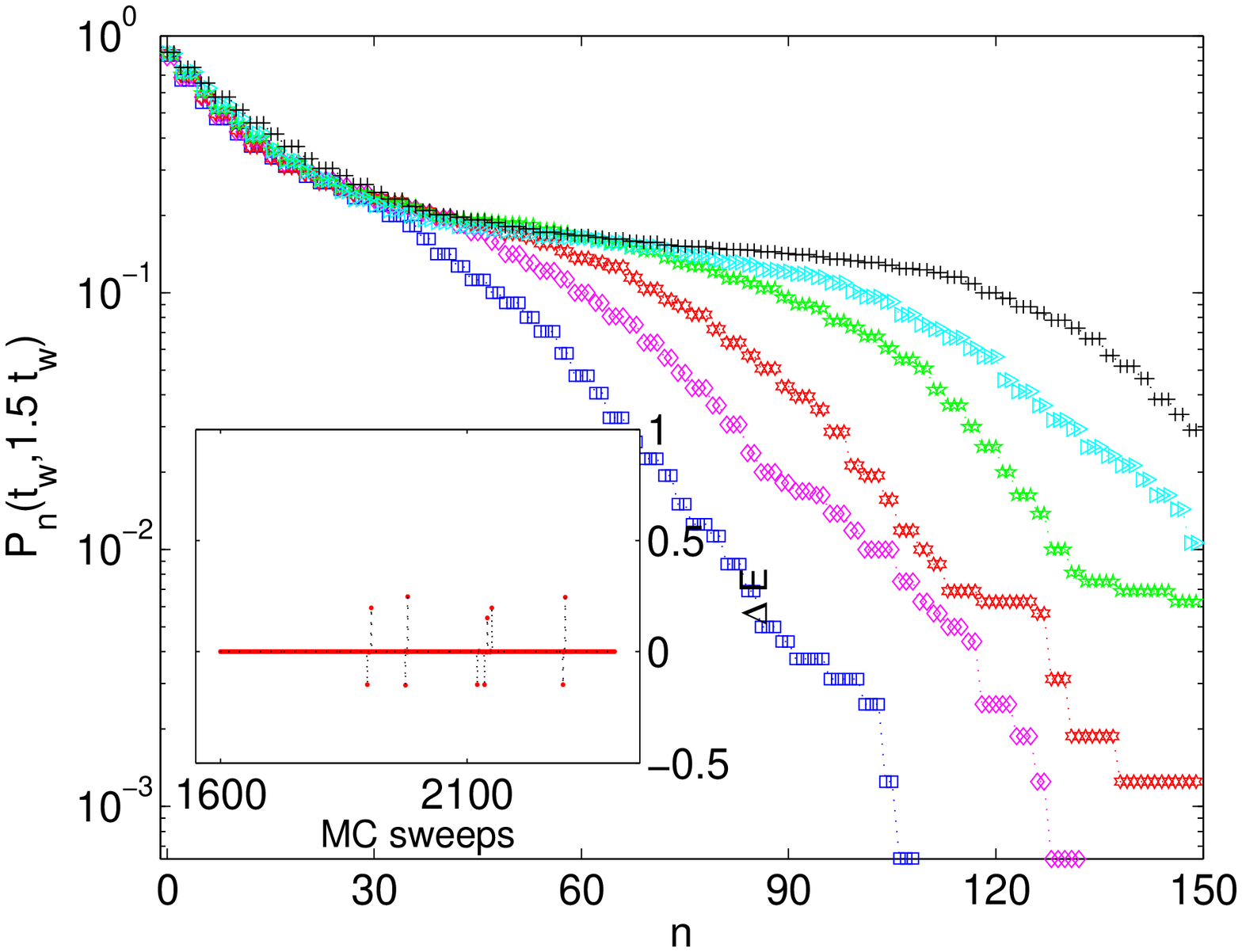}       
\end{array}
$
\caption{(Color online)
 All data shown pertain to a  system with  four ant types,  with one negative and five
positive interactions between different types.
Left panel: the value of the utility function  per ant, averaged over $100$ trajectories, is plotted versus time. The initial configuration
is in all cases obtained by randomly placing the ants on the grid. 
Right panel:  The probability that the number of `large' events (as defined in the main text)
occurring in the interval $[t_w,3/2t_w]$ is larger or equal to   $n$ is estimated using $2500$ trajectories
all run at the DS value  $T=10$.
The results are plotted   versus $n$. for sampling intervals starting at  $t_w=800$ (lowest curve)
up to $t_w=25400$ (highest curve), via intermediate values 
$t_w = 1600,3200, 6400$ and $12800$.
The insert shows a sample trajectory starting at $t_w=1600$. 
}
\label{big_energy_plot} 
\end{figure}

All  data presented in 
Fig. \ref{big_energy_plot} pertain to a  system with four types of ants.
Its  left panel   depicts,  on  a logarithmic
horizontal scale,  the value of the utility  function per ant, averaged over $100$
trajectories, as a function of time (i.e. number of MC sweeps).  For 
$T=500$ and $200$,  a constant value is approached which  increases as the DS
decreases. 
For  $T=5$ and $T=10$  the average  $E$  is seen to grow logarithmically
without  approaching equilibrium. The equilibrium value 
would lie  far above the plateau reached at 
$T=200$. If the logarithmic trend were to continue up to its asymptotic long time limit, the `transient' would 
stretch over more than $10$ decades. The behavior at $T=30$ is  intermediate. 
Here, the curve has two knees, with a third barely visible at the end of the range.
Yet, no stationary state is reached 
within the observation time. 
Consider now the two low $T$ curves.
For  any  fixed value of time, the mean value of the utility function  increases with $T$.
 The strong non-equilibrium character
of the low $T$ dynamics is highlighted by the ordering  being
 opposite than  in the equilibrium regime. 

The statistics of low $T$  fluctuations in $E$  was investigated
for $T=5$ and $T=10$, with data collected at  the latter value of the DS
shown in the right panel of the figure.
The insert illustrates   the  highly intermittent nature of the fluctuations. 
In the main  panel, the probability of large events  $P_n(t_w,\frac{3}{2}t_w]$ is plotted
versus $n$ for six different values of $t_w$, each twice as large as its predecessor.

Data sets belonging to different  observation intervals
collapse  for   values of the abscissa, $n$,  smaller than the length of the observation
interval. Banning finite-time effects, the global dynamics is thus well described by 
a process which is homogeneous in the logarithm of time. This  fully agrees
with the statistical properties of the exit process   which describes  motion at a single site, but which, on the other hand
is observed over a much longer time span. 
We finally note that the same scaling analysis performed using data sampled
at  $T=5$ yields a near
perfect data collapse, with the sole exception of   data   taken during  the shortest 
and earliest observation interval $[800,1200]$

 \section{Summary and outlook}
 
 A  stochastic  model is used to explore  the dynamics
 of groups of  ants, or more generally,  social agents,  which move    in a confined space to   optimize
   a utility function $E$ depending on their  mutual interactions.
  The   dynamics is  stationary, respectively non-stationary  for  high and   low values  of a model parameter   called
  `degree of stochasticity' (DS). The latter is a measure of risk-willingness: at high  DS values, 
  agents readily  perform moves which decrease their utility function, while in the opposite limit they 
  are mainly unwilling to do so.  
 A macroscopically quiescent system state is soon reached at high DS values,
 while the state slowly evolves  due to the gradual emergence of new spatial structures at low values 
 of the DS.
  Correspondingly, the statistics of `exits'  from the nest, i.e. movements involving a particular 
 but arbitrary site,  is   either a simple Poisson
 process ,  as  found by  Nouvellet et al.\cite{Nouvellet10}
or a log-Poisson process as found by Richardson et al.\cite{Richardson10}.
 At low DS  values,  the probability that $n$ large  changes of the utility function occur anywhere in the system
  in a given time interval is found  to   scale  with the difference of the logarithms of the interval's  end-points,
 except for finite  time corrections important for short observation intervals. 
 In conclusion, the model's low DS   dynamics  is
in-homogeneous and decelerating when parameterized by  time, but 
turns into a homogeneous process when the logarithm of time is used as
independent variable.

The above  type of logarithmic relaxation    occurs in  both biological 
and physical systems\cite{Sibani95,Sibani98a,Newman99c,Anderson04,Richardson10}.  
Whenever record-breaking noise fluctuations can trigger irreversible changes
in configuration space its origin is   linked to the mathematical properties of 
record-statistics \cite{Sibani93a,krug05}.
Our model suggests  a novel   interpretation  of record dynamics
as applied to  socially interacting agents: The record sized  fluctuations  triggering
important and irreversible dynamical changes  correspond to  record high investments,
i.e. to a record high  decrease
of  the utility function of the ants involved. 

Returning to the findings of Richardson et al. and Nouvellet et al. which motivated this 
work, one  can speculate on what the present model has to  say on the origin of the
different behaviors  observed by the two groups. Firstly, both behaviors are possible.
Secondly,   the Poisson exit statistics  seen in the  stationary regime
can also more simply  be obtained by assuming that ants act independently of each other.
The log-Poisson behavior does however require interactions. This is the case in 
experiments~\cite{Richardson10a}, where it is removed by disrupting the pattern of social 
interactions among the ants. 
 In the model,  it is possible to increase  the value of the DS parameter in a continuous fashion.
 At a threshold value, the log-Poisson statistics is then replaced by a Poisson statistics.
 The question is whether the same controlled transition could be achieved experimentally,  i.e. by 
 gradually changing  the geometry of the 
 nest, by chemically masking 
 the stimuli ants affect each other with, and finally, by using  drugs which  increase  the 
 risk-willingness of the ants, or equivalently, their indifference to external stimuli.
To obtain the effect might not require  too  drastic interventions: 
A `normal' DS  value just below the threshold could be favored by evolution  because  it 
guarantees  the fastest possible increase of the utility function compatible 
with  the  history dependence and memory effects which generally  occur in glassy dynamics and   
which  seem to naturally  belong to  an evolving  biological organization.
\bibliographystyle{apsrev}
\bibliography{lappe,SD-meld}
\end{document}